# An overview of process model quality literature

## The Comprehensive Process Model Quality Framework

Philipe De Meyer[1], Jan Claes[2*]

[1] *AE nv/sa - Architects for Business & ICT,*
*Interleuvenlaan 27b, 3001 Heverlee, Belgium*

[2] *Ghent University, Department of Business Informatics and Operations Management,*
*Tweekerkenstraat 2, 9000 Gent, Belgium*

\* *Corresponding author. Tel.: +32 9 264 35 17; fax: +32 9 264 42 86.*
*E-mail addresses: philippe.demeyer@ae.be (P. De Meyer), jan.claes@ugent.be (J. Claes).*

**Abstract**

**Purpose** - The rising interest in the construction and the quality of (business) process models resulted in an abundancy of emerged research studies and different findings about process model quality. The lack of overview and the lack of consensus hinder the development of the research field. The research objective is to collect, analyse, structure, and integrate the existing knowledge in a comprehensive framework that strives to find a balance between completeness and relevance without hindering the overview.

**Methodology** - The Systematic Literature Review methodology was applied to collect the relevant studies. Because several studies exist that each partially addresses this research objective, the review was performed at a tertiary level. Based on a critical analysis of the collected papers, a comprehensive, but structured overview of the state of the art in the field was composed.

**Findings** - The existing academic knowledge about process model quality was carefully integrated and structured into the Comprehensive Process Model Quality Framework (CPMQF). The framework summarizes 39 quality dimensions, 21 quality metrics, 28 quality (sub)drivers, 44 (sub)driver metrics, 64 realization initiatives and 15 concrete process model purposes related to 4 types of organizational benefits, as well as the relations between all of these. This overview is thus considered to form a valuable instrument for both researchers and practitioners that are concerned about process model quality.

**Originality** - The framework is the first to address the concept of process model quality in such a comprehensive way.

**Paper type** - Literature review

# 1   Introduction

Business Process Management (BPM) is a discipline that is applied worldwide to improve the productivity of organizations and to help achieve cost reduction (Bandara and Gable, 2012). One of the important activities in the context of the BPM lifecycle is business process modelling. The products of this activity, business process models or process models for short, can be used to "*make decisions about where, how, and why changes to the processes should be enacted to warrant improved operational efficiency*" (Sánchez-González, García, et al., 2013, p. 1) Because process modelling is an activity that is accompanied with many costs (e.g., purchase of a modelling tool, opportunity costs, etc.), the interest in the value proposition of process models is growing (Bandara and Gable, 2012). This value proposition depends on the quality of the resulting process models.

Different studies exist that have contributed to knowledge about the quality of process models (Moreno-Montes de Oca et al., 2014). Some studies focus on the development of theoretical models that support the concept of process model quality (e.g., Krogstie et al., 2006). Other studies focus rather on the development of metrics that characterize specific model properties (e.g., Vanderfeesten, Cardoso and Reijers, 2007). Yet other studies aim to develop practical guidelines to assure or improve the quality of process models (e.g., Mendling et al., 2010). A reported research gap in this domain is that there is no broad overview that brings together these, and potentially other, directions of research about process model quality and that aggregates the built knowledge (Dikici et al., 2017; Recker and Mendling, 2007). However, such an overview is valuable because it forms a solid basis on which new research can build in order to contribute to the development of the knowledge domain (Figl, 2017).

Currently, the research domain is characterized by an abundance of quality dimensions and classifications and by a lack of homogeneity concerning the nomenclature (Sánchez-González, García, et al., 2013). Some authors prefer to approach process model quality from a software perspective (e.g., Vanderfeesten, Cardoso and Reijers, 2007), whereas others make use of quality frameworks for conceptual models (e.g., Sánchez-González, García, et al., 2013). Moreover, for some concepts different terms are used while sharing the same semantics (e.g., comprehensibility and understandability). Although it is possible that authors have different (meticulously built) opinions about these aspects, this lack of consensus causes unambiguity and confusion, which poses a threat to the progress of the domain (Moody, 2005).

In order to improve the transparency in this domain, to aid consensus-building, and to offer material for training and education, this paper describes an *umbrella review* of the literature as defined in (Paré et al., 2015). Therefore, the outcome of the review is nothing less and nothing more than a structured overview of the existing academic knowledge about process model quality. The goal is to be complete, but without endangering the usefulness of the overview.

The collected knowledge is summarized in the "*Comprehensive Process Model Quality Framework*" in Appendix A. This paper is accompanied by a "*data file*" containing all intermediate results http://www.janclaes.info/papers/CPMQF.

Section 2 describes the search and select process to collect the relevant academic literature. Section 3 presents the information resulting from the literature analysis process and it proposes a framework, which visually aggregates these results. To optimize the reading flow of the text, these two sections discuss both the research method and the results of the targeted research phase. Next, Section 4 discusses a number of observations concerning the framework. Finally, Section 5 concludes the paper and identifies the limitations of the study and directions for future research.

## 2 Search and selection of the relevant academic literature

As mentioned before, the research goal is the aggregation and structuring of existing academic knowledge about process model quality.

### 2.1 Research questions

In order to address the research goal, the existing academic knowledge about this topic needs to be collected, analysed, aggregated and structured. Therefore, we defined the next research questions:

- *RQ 1*: Which aspects of process model quality are studied in the literature?
- *RQ 2*: How are the different aspects of process model quality operationalized by literature?
- *RQ 3*: Which structuring efforts are proposed in the literature (i.e., models, frameworks, etc.)?
- *RQ 4*: Which process model quality aspects are accommodated in these proposed structures?

### 2.2 Data collection method

In order to form a scientific answer on these research questions, the available academic literature has to be consulted in a systematic way. As such, the Systematic Literature Review (SLR) is a concrete research method that aims to systematically evaluate and interpret all available literature related to a certain research question. The SLR methodology is assessed as reliable, profound and controllable (Kitchenham and Charters, 2007). In contrast to the conventional SLRs, this study is not primarily concerned with a formal (statistical) comparison of existing knowledge. However, it does go further than a typical mapping review, of which the main goal is to classify the collected studies in an existing framework, because the SLR is used to inductively build such a framework based on a critical assessment of the results. This is similar to the tertiary study presented in (Kitchenham et al., 2010).

Recently, a number of SLRs on process model quality were published (cf. Section 2.2.1 and Section 2.2.2) and other structuring efforts of the domain exist (cf. Section 3.4). However, they address more specific research goals than the broad goal of this paper (cf. Table 1 and Table 2).

Therefore, individually they do not (completely) answer the research questions, and thus we performed a new Systematic Literature Review, albeit at the tertiary level. This is considered to be more efficient and thus more feasible, because the research builds further on the recent structuring efforts and an analysis of a large amount of primary literature is avoided. On the other hand, a critical assumption is that a complete (enough) and accurate view on the research domain can be assembled based on the secondary studies alone.

**2.2.1 Sources and search process**

The literature review was performed as closely as possible to the instructions of Kitchenham and Charters (2007). In order to collect as much relevant literature as possible, a broad automated search strategy in digital databases is preferred. This technique assures that the collection of potentially relevant literature happens in a neutral fashion. Moreover, the selection of four or more databases is considered to raise the probability of completeness (Kitchenham et al., 2010). Inspired by Kitchenham and Charters (2007), we selected the following six databases for the structured literature review: ACM Digital Library, Google Scholar, IEEExplore, ScienceDirect, Scopus, and SpringerLink.

The selected databases are searched based on a simple search string consisting of three parts. The first two parts, "business process model" and "quality", are derived from the research goal. Because of the specificity of "business process model" and the semantic ambiguity of "quality" no synonyms are used for these strings. The third part aims to identify the secondary studies and consists of a series of synonyms of "literature review" defined earlier by Kitchenham et al. (2010). By combining these individual strings with logical operators, next general search phrase was composed:

***Search phrase***: "business process model" AND "quality" AND ("review of studies" OR "structured review" OR "systematic review" OR "literature review" OR "literature analysis" OR "in-depth survey" OR "literature survey" OR "meta-analysis" OR "past studies" OR "subject matter expert" OR "analysis of research" OR "empirical body of knowledge" OR "body of published research")

Based on the fact that the already known secondary studies in the domain (i.e., Moreno-Montes de Oca et al., 2014; Sánchez-González, García, et al., 2013) are recent, we expected the amount of relevant papers to be relatively small. Therefore, a full text search was performed based on this search phrase. The same reasoning supports the decision to also consider the literature that was not published in academic journals or books for this study (such as publicly available master theses), which is also a partial solution to a potential publication bias (the fact that positive results are more likely to be published in academic outlets than negative results).

The Google Scholar database shows only the first 1.000 search results even if more results were found (i.e., 3.862). Therefore, the search string was split in separate strings according to the synonyms of "literature review". Only for the string ("business process model" AND "quality" AND "literature review") still too many results were found (i.e., 2.430). Hence, a number of results could not be collected. Yet, we assume the chance of missing relevant studies to be rather low, because Google Scholar sorts the results according to their interpretation of relevance, and because the search process was applied in multiple databases. An overview of the search process can be found in the *data file*.

### 2.2.2 Selection of studies

In order to assess the relevance of the collected studies for this research, the following selection criteria were applied. They were established according to the research goal and they were refined during the iterative selection process.

The *inclusion criteria* are:

- *IC1*: The principal object of interest of the study has to be (the quality of) the business process model. This should be reflected in the implicit or explicit goal of the central research question.
- *IC2*: The study has to contain conclusions or describe research actions that are related to (a sub-aspect of) the quality of process models.
- *IC3*: The study as a whole has to be a literature review, or at least it should contain a literature review that aims to be representative for the whole body of knowledge in the targeted aspect of quality at the time of its publication.

The *exclusion criteria* are:

- *EC1*: Studies that focus on topics of the applied domains of for example project management, healthcare management and education management are excluded.
- *EC2*: Studies that investigate the quality of process mining, process simulation or process languages in particular are excluded.
- *EC3*: Studies that investigate the efficiency of the process of process modelling without contributing to knowledge about the quality of the end product (the model) are excluded.

Finally, two *practical limitations* manipulated the selection process:

- *PL1*: The full text has to be available in public or through the subscriptions of Ghent University.
- *PL2*: Studies in other languages than English are excluded.

The selection process proceeded in two phases. First, the search results of the automatic search process were analysed to identify the relevant studies. Second, snowballing was applied, which is a technique of literature reviews that considers the references of the identified studies of the first phase in order to select additional relevant studies if necessary. This two-phase approach is consistent with the recommendation of Kitchenham and Charters (2007) to not just rely on the digital search strategy. The manual search of references improves the representativeness of the literature review.

The first selection phase started with the application of the selection criteria on the title, abstract and keywords, resulting in a set of 121 studies. It is important to note that in case of doubt the papers were not discarded in this step. The next step is the application of the selection criteria on the full text, providing an initial set of 24 studies. For each paper, the selection criteria were assessed by only one of the authors, whereas for one arbitrary year (i.e., 2014) the papers were analysed by both authors. No differences were found in their classification, meaning that both authors independently selected the same four papers based on their interpretation and assessment of the selection criteria.

In the second selection phase, the 1.449 references from the 24 studies were processed in the same manner, based on the same selection criteria. As such, a long list was composed of 118 studies, which was reduced in the next step to 18 relevant studies. The final set thus contains 42 relevant studies. An overview of the search and selection process is presented in Fig. 1.

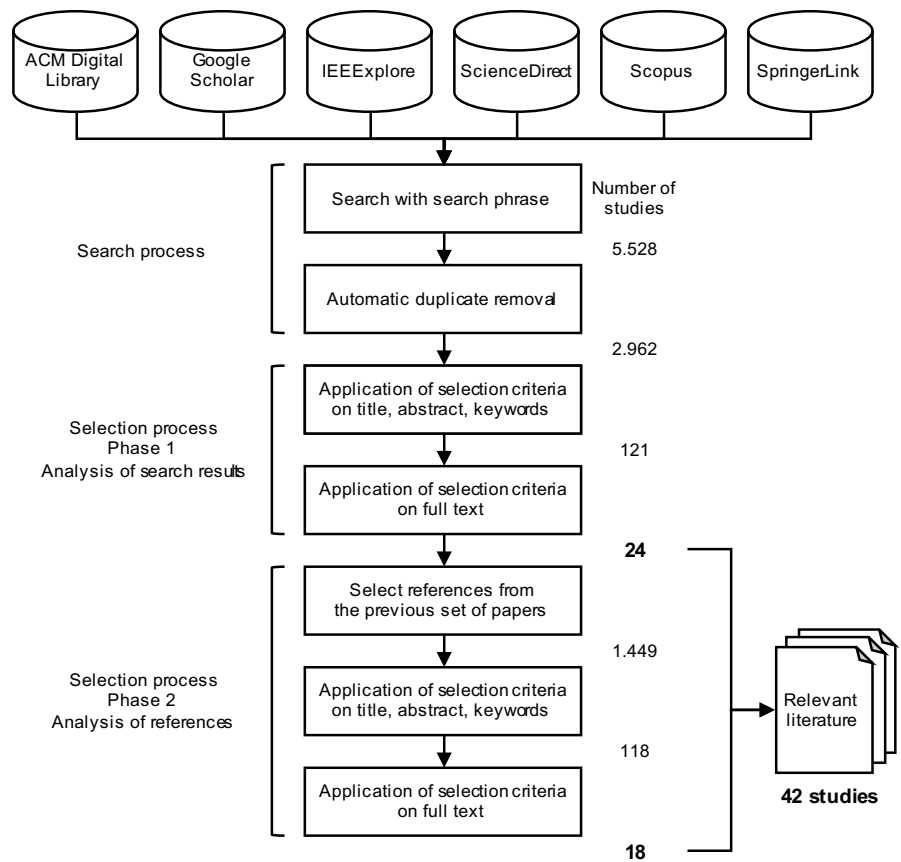

Fig. 1. Search and selection process

### 2.2.3 Assessment of the quality of the secondary studies

The absence of publication or review requirements as selection criteria may lead to the inclusion of studies of low quality or low reliability. Hence, according to the guidelines of Kitchenham and Charters (2007), the DARE criteria (http://www.york.ac.uk/crd/#DARE) were applied to assess the quality of the collected studies. The criteria are based on four questions:

- *C1*: Does the literature review describe and apply inclusion and exclusion criteria?
- *C2*: Is it probably that the literature review considers all relevant studies?
- *C3*: Do the researchers assess the quality or validity of the studies?
- *C4*: Were the source data and studies described adequately?

These questions are coded as follows (Kitchenham and Charters, 2007):

- *C1*: Y (yes), the inclusion criteria were defined explicitly in the paper, P (partial), the inclusion criteria are implicit, N (no), the inclusion criteria are not defined and cannot be derived.
- *C2*: Y, the authors searched 4 or more digital libraries and applied additional search strategies, or they identified all journals that target the selected research topic, P, the authors searched 3 or 4 digital libraries and did not apply additional search strategies, or they searched a well-defined but limited set of journals, N, the authors searched 2 or less databases or an extremely small set.
- *C3*: Y, the authors have defined and evaluated the quality criteria explicitly for each primary study, P, the research question takes quality requirements for the literature review into account, N, no specific actions are described that assess the quality of individual papers.
- *C4*: Y, information about each paper is presented, P, only summarized information is presented about each individual paper, N, the results of individual papers are not specified.

Kitchenham and Charters describe in their guidelines that researchers should focus in their quality assessment on the evaluation of the research method and not of the reporting quality. Not all aspects of a research method are (equally detailed) described by every researcher, yet the absence of the reporting does not necessarily imply the absence of the method. Nevertheless, we implemented a conservative approach for the evaluation of quality. The absence of information in the report about a particular criterion has led to a negative evaluation of the criterion (N), causing the risk that the quality of some papers was estimated too low.

Table 1 provides a chronologic overview of the assessment of the 42 selected studies on the different quality criteria. From this overview, it can be concluded that only 5 papers score well on all defined quality criteria (C1-C4). These literature reviews seem to be methodologically strong and are thus classified as real Systematic Literature Reviews. Some authors claim to make an effort to execute their literature review in a systematic way (i.e., Braunnagel et al., 2014; Van Mersbergen, 2013), but they failed to report this accurately (enough) to get a good quality score on all criteria.

A majority of papers are falling short concerning quality criteria C2 and C3. These are secondary studies that have no guarantee to be sufficiently complete, and that did not evaluate the quality or validity of the primary studies. Nevertheless, it is possible that these studies tackle other aspects than the five identified Systematic Literature Reviews, even if they may individually be not complete or describe information of low quality. Their inclusion improves the completeness but can obviously introduce lower quality information in the overview. Furthermore, the decision to not only include literature published in academic outlets seems to have a positive outcome. It has enabled the inclusion of for example the master thesis of Sadowska (2013), which has extended the limited set of Systematic Literature Reviews. Its inclusion has improved the internal validity of this tertiary study.

Table 1. Quality assessment (C1-C4) and content classification (Q1-Q6) of the selected secondary studies

| Study | C1 | C2 | C3 | C4 | Q1 | Q2 | Q3 | Q4 | Q5 | Q6 |
|---|---|---|---|---|---|---|---|---|---|---|
| (Rosemann et al., 2001) | P | N | N | Y | 1 | | | | 1 | |
| (Gruhn and Laue, 2006) | P | N | N | Y | | | | 1 | | |
| (Krogstie et al., 2006) | P | N | N | Y | 1 | | | | | |
| (Recker, 2006) | P | N | N | Y | 1 | | | | | |
| (Mendling, 2007) | P | N | N | Y | | 1 | 1 | 1 | | |
| (Recker and Mendling, 2007) | P | N | P | Y | 1 | 1 | | | | 1 |
| (Vanderfeesten, Cardoso, Mendling, et al., 2007) | P | N | N | Y | | | 1 | 1 | | |
| (Vanderfeesten, Cardoso and Reijers, 2007) | P | N | N | Y | | | 1 | 1 | | |
| (Azim et al., 2008) | P | N | N | Y | | | 1 | 1 | | |
| (Mendling and Strembeck, 2008) | P | N | P | Y | 1 | | 1 | 1 | | |
| (Khlif et al., 2009) | P | N | N | Y | | | | 1 | | |
| (Matook and Indulska, 2009) | P | N | P | Y | 1 | | | | | 1 |
| (Cappelli et al., 2010) | P | N | N | Y | | | | 1 | | |
| (Khlif et al., 2010) | P | N | N | Y | | | | 1 | | |
| (Laue and Mendling, 2010) | P | N | N | Y | 1 | 1 | 1 | | | |
| (Muketha et al., 2010) | Y | N | N | Y | | | | 1 | | |
| (Reijers et al., 2010) | P | N | N | Y | | | 1 | 1 | | |
| (Sánchez-González et al., 2010) | Y | Y | Y | Y | 1 | | | 1 | | |
| (La Rosa, Ter Hofstede, et al., 2011) | P | N | N | Y | 1 | | | | 1 | |
| (La Rosa, Wohed, et al., 2011) | P | N | N | Y | 1 | | | 1 | 1 | |
| (Laue and Gadatsch, 2011) | P | N | N | Y | | 1 | | | | |
| (Reijers and Mendling, 2011) | P | N | N | Y | | | 1 | 1 | | |
| (Bandara and Gable, 2012) | N | N | N | Y | 1 | | | 1 | | |
| (Dumas et al., 2012) | P | N | N | Y | | | | 1 | | |
| (Houy et al., 2012) | N | P | P | Y | | 1 | | | | |
| (Krogstie, 2012) | N | N | N | Y | 1 | | 1 | | 1 | |
| (Mendling et al., 2012) | Y | N | N | Y | | | 1 | 1 | 1 | |
| (Sánchez-González et al., 2012) | P | N | N | Y | | | 1 | 1 | 1 | |
| (Soffer et al., 2012) | P | N | N | Y | | | 1 | | | |
| (Becker et al., 2013) | P | N | N | Y | 1 | | | 1 | | |
| (Sadowska, 2013) | Y | Y | Y | Y | 1 | | | 1 | 1 | 1 |
| (Sánchez-González, García, et al., 2013) | Y | Y | Y | Y | 1 | | 1 | 1 | 1 | |
| (Sánchez-González, Ruiz, et al., 2013) | P | N | N | Y | 1 | | 1 | 1 | | |
| (Van Mersbergen, 2013) | P | N | Y | Y | 1 | 1 | 1 | 1 | 1 | |
| (Braunnagel et al., 2014) | P | P | N | Y | | | 1 | 1 | | |
| (Houy et al., 2014) | Y | N | N | Y | | | | | | |
| (Kluza et al., 2014) | P | N | N | Y | | | 1 | 1 | 1 | |
| (Moreno-Montes de Oca et al., 2014) | Y | Y | Y | Y | 1 | | | | 1 | |
| (Reijers et al., 2015) | P | N | P | Y | 1 | 1 | | | 1 | |
| (Rompen, 2015) | N | N | N | Y | | | 1 | | | |
| (Polančič and Cegnar, 2016) | Y | Y | P | Y | | | 1 | | | |
| (Figl, 2017) | Y | Y | Y | Y | | 1 | 1 | 1 | | |
| Total | | | | | 19 | 9 | 22 | 22 | 13 | 3 |

# 3    Analysis and structuring of the collected literature

Below, the collected knowledge from the secondary studies is discussed per research question.

## 3.1   Identification of the aspects of process model quality (RQ 1)

To formulate an answer to the first research question, it is important to identify a classification criterion to split the paper set in sub-sets. This way a reference framework emerges that makes it easier to situate and answer the next research questions. Inspiration for such a classification was found in the field of *Enterprise Architecture* (EA). An influential work in this field is the 'Zachman Framework For Enterprise Architecture' (Zachman, 2003), which presents a multidimensional classification scheme for architectural artefacts. The described objectives are similar to the objectives of this paper: (1) provide insights in the EA domain (here: the domain of process model quality), and (2) organize and position the different EA artefacts (here: the concepts of process model quality).

Zachman organizes the framework according to six kinds of abstractions, which can in essence be captured by six questions (i.e., what, how, why, where, who, and when). These questions are primitive (i.e., they each address a different aspect) and comprehensive (i.e., together, they address all relevant aspects). However, during the selection phase the impression arose that the collected studies do not discuss aspects related to the 'where' and 'when' of process model quality. Moreover, the 'who' question appears to be always treated as a specialization of the 'what' or 'how' question in the identified studies, which makes it no longer a primitive question in such cases. Although the 'where', 'when' and 'who' questions are conceptually still relevant and yield potentially valuable research insights, the three other questions are deemed most useful to classify the described knowledge. The interrogatives can be further specified in key questions that aim to be representative for the different aspects of process model quality. They jointly address the *research goals* that are defined by Genero et al. (2011) and that are used as classification instrument by Moreno-Montes de Oca et al. (2014).

The six key questions are:

- *Q1*: WHAT is process model quality?
- *Q2*: HOW is process model quality measured?
- *Q3*: WHAT are the drivers of process model quality?
- *Q4*: HOW are the drivers of process model quality measured?
- *Q5*: HOW is process model quality realized?
- *Q6*: WHY is process model quality pursued?

It can be noticed how some of these questions (i.e., Q1, Q3, Q6) have a strong focus on deriving a more abstract/theoretical insight in the domain, whereas the other questions (i.e., Q2, Q4, Q5) are aimed at the development of concrete/practical methods and guidelines. As shown further in this paper, this research domain contains continuous efforts to bridge theoretical and practical aspects. Each of the selected studies provides, explicitly or not, answers to one or more of the above defined key questions. Table 1 also summarizes our classification of papers based on their content. As such, it offers a first useful insight in the content of the paper set.

### 3.2 Description of the aspects of process model quality (RQ 2)

The six identified key aspects of process model quality are discussed below to address RQ 2. Their relation is presented in Fig. 2. Process model quality (Q1) can be considered as a variable that depends on a number of drivers (Q3). A distinction is made between the theoretical definition of these concepts (Q1 and Q3) and their respective measurement (Q2 and Q4). Furthermore, initiatives exist that operationalize process model quality (Q5) by providing guidelines to influence the drivers of process model quality. The literature also motivates the relevance of process model quality (Q6).

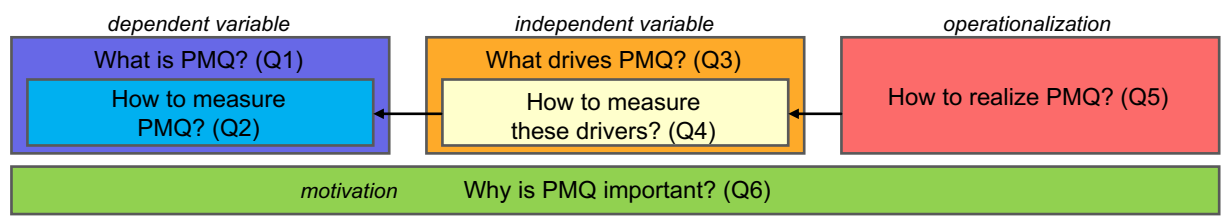

Fig. 2. Relation between the 6 key questions that represent the different aspects of process model quality (PMQ)

### 3.2.1 WHAT is process model quality (RQ 2 - Q1)?

This question is related to the general concept of quality. Garvin (1984) identified five approaches to define the concept of quality. First, there is the *transcendent approach*. It states that quality is an "innate excellence", which is absolute and universally recognizable yet not exactly definable. Second, the *product-based approach* defines quality as a precise and measurable variable influenced by the presence of a product attribute in a specific quantity. In contrast, the *user-based approach* states that quality is the degree to which the preferences of a consumer are satisfied, which is a strong subjective property. There is also the *manufacturing-based approach*. It describes quality as the degree to which a product meets the requirements from the design phase. Finally, in the *value-based approach* a product or service has high quality if it is performant at an acceptable price or cost.

As Garvin correctly points out, these different perspectives can lead to miscommunication and conflicts within and between companies. Hence, Garvin proposes a cascade in which the preferences are identified by market research (user-based approach), are then translated into product attributes (product-based approach), to which the production process is aligned (production-based approach). Similarly, the process model quality literature conforms to the product-based (cf. Q2 and Q4) and the production-based approach (cf. Q5), while not disregarding the user-based approach (cf. Q6). Indeed, a well-known definition of quality, is the ***fit-for-purpose*** (Juran et al., 1974). Process models cannot have a high quality if they are not suitable for achieving their goal(s), their *'raison d'être'*.

In recent decades, an abundance of models have been developed to theoretically support the notion of (process) model quality (Nelson et al., 2012). Below, we briefly discuss a selection of the most occurring models in literature, which are also considered the most relevant. To not compromise the usability, only these models are included in the final overview (cf. Fig. 3). The SEQUAL framework and its extensions are not included separately, because they are incorporated in the CMQF framework. The *data file* presents all the identified quality frameworks and standards.

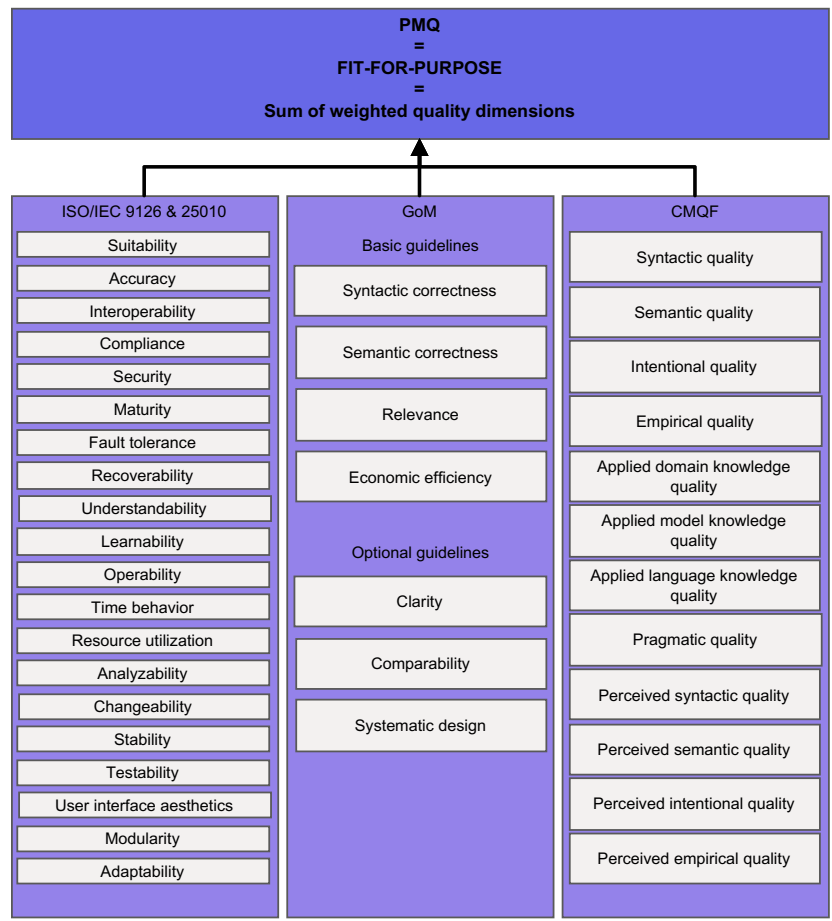

Fig. 3. Types of process model quality

**The SEQUAL framework**. One of the oldest and most influential frameworks about the quality of conceptual modelling is the SEQUAL (semiotic quality) or LSS-framework (Lindland et al., 1994). No more than 14 of the 19 papers discussing Q1 refer to this work (73,68%). The framework is based on semiotic theory (Morris, 1938). By considering the conceptual models as sets of statements in a language, the evaluation in linguistic and semiotic terms becomes possible (Krogstie et al., 2006). In its original form, the framework proposes *3 types of quality*:

- Syntactic quality: the symbol accordance with the modelling language syntax and vocabulary.
- Semantic quality: the correctness and completeness of the model in relation to reality.
- Pragmatic quality: the understanding correctness of the model by its users.

**The SEQUAL extensions**. The SEQUAL framework has been adapted and extended multiple times, e.g. by Krogstie et al. (2006), who make a distinction between *10 types of quality*. An important limitation of the SEQUAL framework and the extensions is the static perception of the domain. We propose that the model can be a conductor for change in the domain (a to-be process model). Therefore, a distinction should be made between the factual domain and the optimal domain in order to describe the semantic quality for '*as is*' and for '*to be*' process models in the proper terminology.

**The CMQF framework**. A more recent effort which builds further on the aforementioned frameworks and which was identified 3 times in the 19 papers, is the CMQF (Conceptual Modelling Quality Framework) (Nelson et al., 2012). This framework synthesizes the above-mentioned SEQUAL extension and the Bunge-Wand-Weber (BWW) framework (Wand and Weber, 1990). The latter takes a process-oriented point of view on quality, in contrast to the product-oriented viewpoint of the SEQUAL framework (Nelson et al., 2012). The CMQF makes a distinction between the physical (*real world*) and the social reality (*in the mind*). Since the focus is on the quality of representations (i.e., the quality of process models), it makes sense to only consider these types of quality that relate to the *physical representation* (the process model) or the *representation knowledge* (the interpretation of the model). The CMQF defines *12 types of representation quality*.

**The GoM framework**. A framework that centres specifically on process model quality and which is grounded in general accounting principles, is GoM (*Guidelines of Modelling*) (Becker et al., 2000). It is referred in 11 of the 19 relevant studies and it bundles 6 general principles that influence the quality of a model, divided into essential *basic guidelines* and desired *optional guidelines*. Some critical concerns and comments can be raised for this framework. First, the distinction between *basic* and *optional guidelines* is debatable. What is essential and what is optional, depends heavily on the purpose of the process model. For example, if a model is created to support communication about a process, it is not unreasonable to assume that *clarity* is an essential dimension. Second, the framework lacks the theoretical and methodological grounds of the formerly mentioned frameworks. Therefore, the proposed guidelines appear to be a random set and they are not unambiguously defined (e.g., the *clarity* guideline mixes up the concepts of *readability* and *understanding*). Third, it proposes guidelines about the desired properties of a process model, but it does not offer the methods to reach such an outcome. The framework is thus merely a theoretical knowledge contribution and the choice of the term *guidelines* seems to be rather unfortunate.

**The ISO/IEC quality standards**. Many authors make use of the similarities between software applications and process models for their research towards the different aspects of quality (e.g., Gruhn and Laue, 2006). Vanderfeesten et al. (2007) summarized three similarities that are frequently mentioned in literature: (1) both domains focus on information processing: inputs are transferred into outputs, (2) they have a similar composing structure, and (3) the dynamic execution is derived from a static structure. The application of the ISO standards for the quality of software on process models is mentioned in 5 of the 19 studies that address Q1. Sánchez-González, et al. (2013) compared the ISO/IEC 9126 and ISO/IEC 25010 standards and selected the most relevant characteristics, completed with a number of external characteristics (cf. *data file*). To compare process models with software applications is beneficial because the well documented and generally accepted ISO standards can be applied, which is valuable in a domain that is characterized by a lack of consensus. Although process models and software applications are similar in their functioning and structure, it should not be forgotten that both are different entities with different goals. As is expressed by Recker and Mendling (2007), the facilitation of communication is one of the key objectives of process models. It is doubtful whether this objective is equally crucial for software applications. The ISO standards therefore potentially do not capture all relevant characteristics, which explains why additional external characteristics are included by Sánchez-González, et al. (2013). Moreover, in order to improve the understandability for the application of process models, the ISO characteristics are in most cases adapted (Sánchez-González, García, et al., 2013), which hinders their direct application.

### 3.2.2 HOW is process model quality measured (RQ 2 - Q2)?

This question targets the identification of qualitative or quantitative measurements for the discovered dimensions. As can be observed in Table 1, the number of identified studies that investigate this question is rather limited. There are two potential explanations. First, as determined above, there is a lack of consensus about which dimensions constitute process model quality. Hence, is it cumbersome to decide how a certain dimension should be measured and which dimensions are relevant to be measured. Second, many quality dimensions appear to be abstract by nature and/or it is practically impossible to measure them objectively because a personal assessment needs to be made. An analysis of the paper set (cf. *data file*) reveals that mainly syntactic quality/correctness is investigated as a measurable quality dimension, whereas far less papers investigate the measurement of the pragmatic quality/understandability and also the modifiability of process models (cf. Fig. 4).

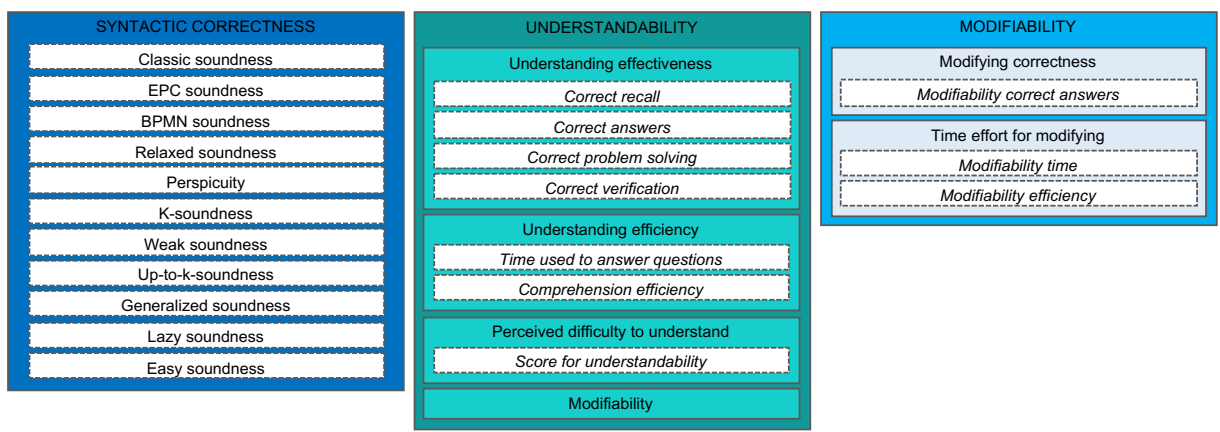

Fig. 4. Allocation of metrics to process model quality dimensions

**Measurement of 'syntactic correctness'.** The term syntactic correctness is used by Mendling (2007) and by Van Mersbergen (2013) and is clear in the sense that it avoids confusion with the term semantic correctness. The literature identifies one main correctness criterion that focuses on avoiding grammatical irregularities such as deadlocks and live locks (Laue and Mendling, 2010; Recker and Mendling, 2007; Reijers et al., 2015). This criterion is known as the soundness property (Van der Aalst, 1997) and comprises three requirements in order to assure the correct execution of a workflow-net with one source and one sink: (1) from each state, it is possible to reach the end state, (2) each end state is a full end state, no active branches are left, and (3) there are no branches in the model that can never be executed. With the aid of tools, it can be automatically assessed if these requirements are met. Also, some tools are programmed in such a way that they do not allow a user to perform syntactically wrong operations (e.g., ARIS Express BPMN does not allow to introduce a message flow within a BPMN swim lane).

**Measurement of 'understandability'.** The term understandability is preferred by Laue and Gadatsch (2011) and is considered to be a valid synonym for the term comprehensibility preferred by Van Mersbergen (2013). It is important to note that, in contrast to what the various authors appear to suggest, the understandability of a model actually cannot be measured directly. As is posed by Krogstie et al. (2006, p. 94), "*it is really only syntactic quality (...), which has any hope of being* [directly] *objectively measured, as both the problem domain and the minds of the stakeholders are unavailable for formal inspection*".

**Measurement of 'modifiability'.** The term modifiability is used in different papers (Sánchez-González et al., 2012; Sánchez-González, García, et al., 2013) and can be considered as a synonym of changeability and as sub-dimension of the concept maintainability (cf. ISO/SEC 25010). Sánchez-González et al. (2012) described an experiment of which they claim that it measures modifiability of a model as a dependent variable. However, again we argue that modifiability cannot be directly measured, but it can only be approximated with proxies.

### 3.2.3 WHAT are the drivers of process model quality (RQ 2 - Q3)?

This question aims to reveal the factors that influence the composing drivers of process model quality. In the paper set, 22 literature studies have been identified that answer Q3 (cf. Table 1). The *data file* provides an overview of the discovered drivers and their related quality dimensions. Two observations are made. First, the list of identified drivers is extensive and their nomenclature is inconsistent (e.g., modularity and modularization), as is their structure (e.g., size is mentioned as an autonomous driver, but also as a determinant for complexity). This supports the observation of a lack of general consensus in the research domain about terminology and classification. Second, there is an abundance of research about the drivers that relate to understandability and to a lesser extent maintainability. This is in accordance with the observations of Moreno-Montes de Oca et al. (2014).

In order to keep an overview, we applied an initial structuring of the findings (cf. Fig. 5), based on the categories proposed by Mendling (2007) and by Reijers and Mendling (2011). Model-related factors have a crucial role. They are subdivided in factors that relate to the abstract (structure) or the concrete syntax (graphical layout) of the models. Personal factors were also found, as well as a small number of other, rather context-related factors.

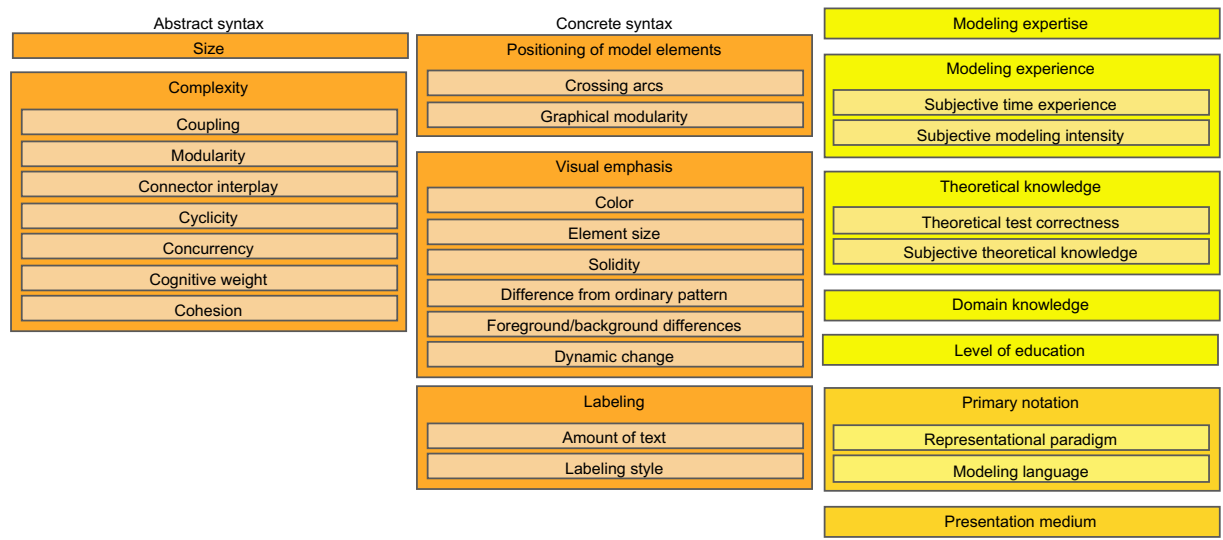

Fig. 5. Drivers of process model quality

**Model-related factors**. The majority of the selected studies discuss model-related factors that relate to the abstract syntax of a process model. This concept describes the structure of process elements and their mutual relations (La Rosa, Wohed, et al., 2011). There is also a noteworthy role for model-related factors that relate to the concrete syntax of process models. The focus here is on the graphical representation of the models.

**Personal factors**. Reijers and Mendling (2011) investigated the factors that influence the understandability of process models. In their work, the authors direct attention to the important role of personal factors, because understandability implies a human interpretation.

**Other factors**. The literature review shows that the majority of the research focuses on personal and model-related factors as drivers of process model quality. Yet, some authors identify other drivers (Mendling and Strembeck, 2008; Soffer et al., 2012), such as modelling language and representation medium.

### 3.2.4 HOW are the drivers of process model quality measured (RQ 2 - Q4)?

This key question attempts to reveal qualitative or quantitative measurements for the drivers of process model quality. As can be seen in Table 1, of all the key questions (Q1-Q6) this question is considered the most in the collected literature. It can be concluded that the research into the driver metrics is a *major topic* in the research domain.

**Model-related metrics**. In order to create an overview of the metrics that are based on abstract syntax, the paper set is investigated and per study we identified the described metrics and the drivers they are proposed to measure (cf. *data file*). Exactly 118 unique metrics were found with varying frequencies (i.e., the number of studies that mention the metric). The sum of the frequencies of all these metrics is 391. However, not all 118 metrics are incorporated in the final framework, because this would hinder the usefulness of the overview that this paper aims to create (cf. Fig. 6). Hence, only those metrics that were mentioned in at least 3 papers are discussed. This resulted in a selection of 40 unique metrics representing 305 of the 391 mentions (78%).

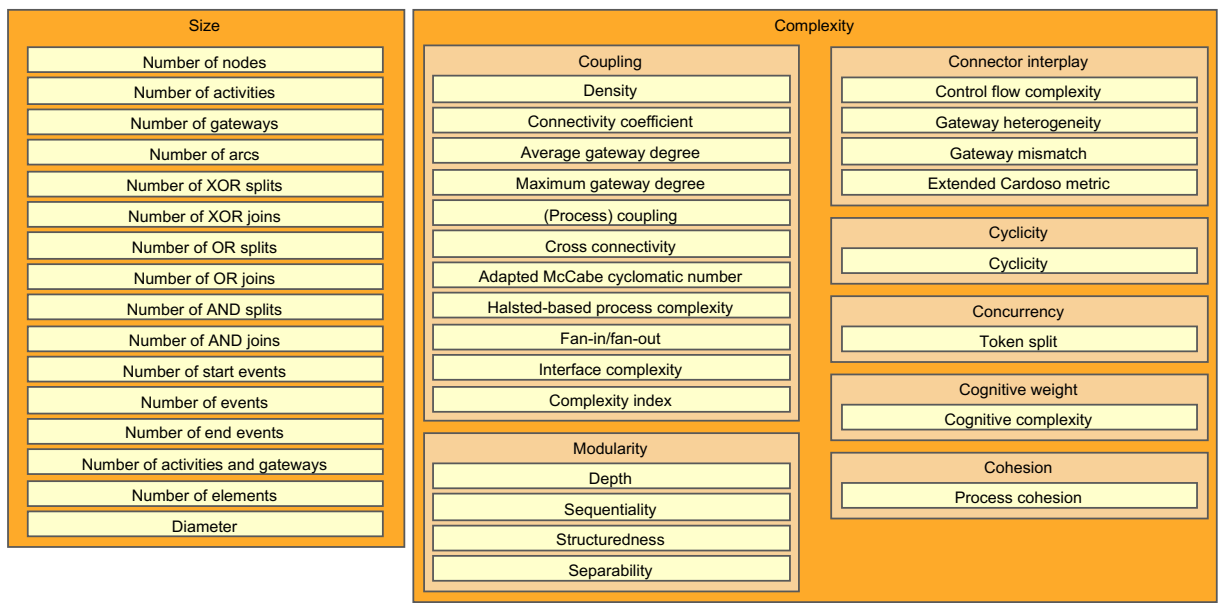

Fig. 6. Allocation of model-related metrics to the process model quality drivers

**Personal metrics**. Fewer papers were identified that discuss the measurement of personal factors influencing the process model (cf. Fig. 7). In contrast to the model-related factors, these factors are not directly and objectively measurable, because formal inspection of the thoughts or of the full lifespan of the test subjects is infeasible.

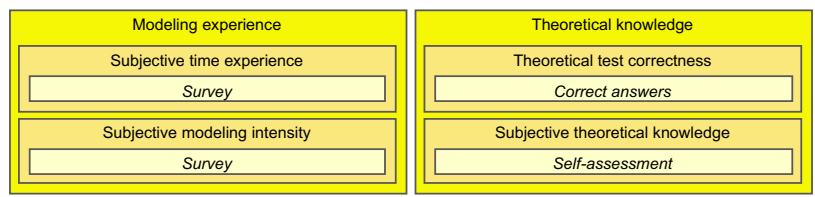

Fig. 7. Allocation of personal metrics to the process model quality drivers

### 3.2.5 HOW is process model quality realized (RQ 2 - Q5)?

This question focuses on the identification of initiatives that aim at assuring process model quality or at improving process model quality. Such initiatives for realizing a high degree of process model quality are unquestionably of great practical value for professional modellers. A detailed overview can be found in the *data file* and the initiatives included in the framework are summarized in Fig. 8.

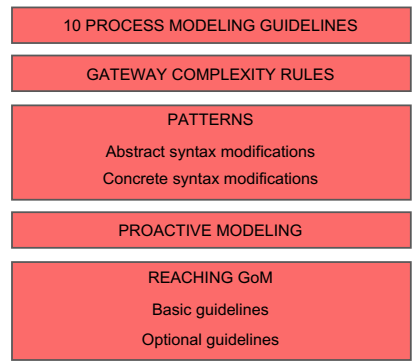

Fig. 8. Initiatives to realize process model quality

**Guidelines**. The most observed instrument for the realization of process model quality is the collection of guidelines termed 7PMG (*Seven Process Modelling Guidelines*) (Mendling et al., 2010). These guidelines are based on empirical research and are supposed to guide the modeller in constructing models that are understandable and that have a lower chance to contain syntactic errors. The 7PMG do not form a restrictive set of do's and don'ts which always must be met. They are considered as desirable modelling practices that generally produce good results. Their empirical foundation makes them more reliable than personal modelling preferences (Mendling et al., 2010). For a set of metrics, Mendling et al. (2012) determined thresholds for the prediction of syntactic errors in process models. They can be used to guide the modeller during the design phase. Based on these thresholds, the 7PMG were refined and extended to 10PMG (*Ten Process Modelling Guidelines*).

**Gateway complexity rules**. Sánchez-González et al. (2012) stress the importance of metrics as objective process model quality indicators. They attempted to relate a value interpretation to a number of gateway complexity metrics. They used statistical techniques on data from a controlled experiment to determine thresholds that the modeller should monitor and should try not to exceed in order to achieve the predetermined degree of understandability and modifiability. From these thresholds, the authors distil a number of concrete guidelines for process modelling.

**Patterns**. In 2011, two studies were published that propose patterns to control the complexity of process models via concrete (La Rosa, Ter Hofstede, et al., 2011) and abstract (La Rosa, Wohed, et al., 2011) syntax modifications. The papers assume that the application of the patterns influences certain model-related metrics. Obviously, none of the patterns guarantee a high process model quality when applied in extreme. For example, the addition of icons can increase the understandability of models, but when the number of icon types becomes too high, the model will be less understood after all. Moreover, some patterns contradict (such as 'duplication' and 'compacting') and thus choices need to be made to balance the opposing effects of different patterns on the process model quality.

**Proactive modelling**. One of the existing quality frameworks identified in this literature study, is the SIQ framework (Reijers et al., 2015). It includes techniques that assure process model quality.

More specific, it aims at protecting the syntactic quality (*correctness-by-design*), the semantic quality (*truthfulness-by-design*) and the pragmatic quality (*understandability-by-design*).

**Reaching GoM**. In a relatively early paper of Rosemann et al. (2001) initiatives to accord to GoM (*Guidelines of Modelling*) are identified based on a literature review. In contrast to the ones discussed above, these initiatives are not very concrete. Moreover, they do not specifically focus on model characteristics of well-defined modelling methods, but they focus mainly on elements of the modelling context. As the authors propose in the conclusion of their study, the relative importance of these initiatives strongly depends on the goal of the process model. This is in line with the general view on process model quality as '*fit-for-purpose*'.

### 3.2.6 WHY is process model quality pursued (RQ 2 - Q6)?

At the beginning of this section, we proposed to judge the general quality of process models as the degree of fit-for-purpose. The pursuit for models of high quality is thus meaningful, because they are per definition capable to realize their goals. Depending on these goals, different relative importance should be attached to the specific dimensions. This brings us to the question which are the different goals of process models (Q6), according to the insights of the selected literature (cf. Fig. 9).

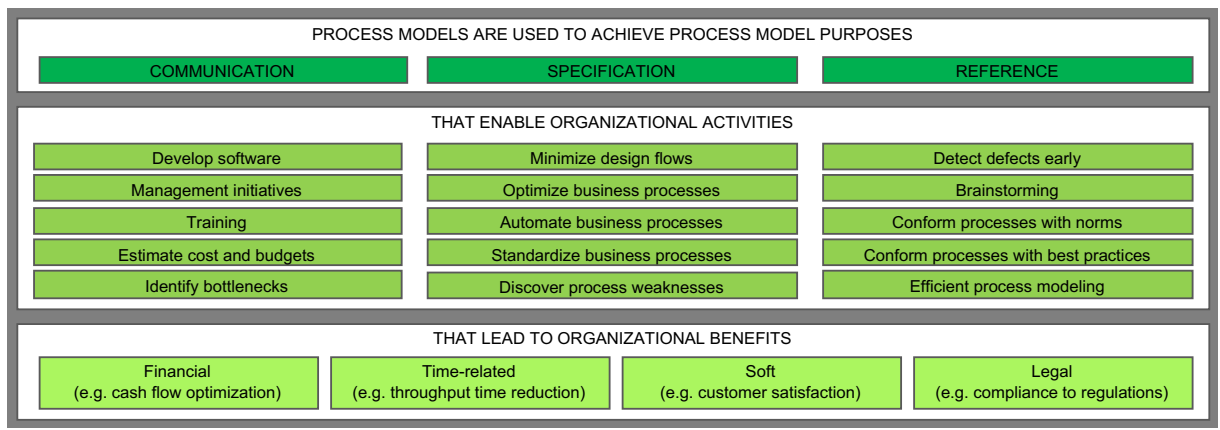

Fig. 9. Goals of process models

The analysis of the literature revealed a number of goals for process modelling: i.e., communication (Recker and Mendling, 2007), specification (Recker and Mendling, 2007) and reference (Matook and Indulska, 2009). These goals can be interpreted as services that could be delivered with the process models. These services support in turn a variety of business activities in order to realize specific organizational benefits. Matook and Indulska (2009) and Sadowska (2013) list a number of such business activities and organizational benefits (e.g., customer satisfaction). To produce an exhaustive or even extensive list of those organizational benefits is out of the scope of this paper. Yet, as an illustration, we included four benefits in the framework: financial benefits (e.g., cash flow optimization), time-related benefits (e.g., throughput time reduction), more 'soft' benefits (e.g., customer satisfaction) and legal benefits (e.g., compliance to various regulations).

It can be concluded that process models of high quality indirectly contribute to achieving specific organizational benefits. Note that also the opposite holds. Low quality process models may insufficiently support business activities, in which case the benefits may not be realized. In the worst case, low quality models can even harm the organization: for example, through a systematic miscommunication caused by a weak semantic quality or through the excessive maintenance costs of models with a low modifiability. Therefore, it is important that the quality of the utilized process models in organizations is high (enough).

### 3.3 Aggregation of the collected literature in a comprehensive framework

The knowledge presented in Section 3.2 is bundled and structured in a comprehensive process model quality framework, presented in Appendix A (a high-resolution, printable version can be downloaded from www.janclaes.info/CPMQF). We are convinced that the complete framework presents a comprehensive and structured view on the current state of the process model quality research domain. For this reason, it is called the Comprehensive Process Model Quality Framework (CPMQF). As can be derived from the explanation below, it is gradually built based on the six key questions (Q1-Q6). The concept of process model quality is abbreviated as 'PMQ'.

The link between the quality dimensions (Q2) and the quality frameworks and standards (Q1) is not always obvious. Although, it seems clear that syntactic correctness relates to syntactic correctness from GoM and syntactic quality from CMQF, the link with understandability and with modifiability is less clear. The different proxies for both dimensions appear to sometimes address different quality dimensions per framework.

Then, after linking metrics (Q4) to the corresponding drivers (Q3), the relationships between the drivers and the PMQ dimensions (Q2) were investigated. The definition of the drivers and the metrics was used to guide this activity and our findings were than compared to the collected literature. Whenever the literature mentions a validated and significant relation between two concepts, they are connected in the framework with a solid line. If applicable the line is annotated with a '+' or '-', which represents the direction of the described correlation. In case of conflicting reported correlation directions, the directions that was reported most was preserved. An overview of the information is presented in the *data file*. It should be noted that only significant results were considered, whereas we also found many studies that showed certain relations to be insignificant.

Finally, the initiatives to improve PMQ (Q5) are linked to the drivers that they impact (according to the respective authors). When a study reveals a link between an initiative and a certain driver or corresponding metric, they are connected in our framework with a dashed line towards the affected driver or metric. When the relation is mentioned to be positive (the initiative improves or protects the driver or metric), this is denoted with a '+'. The opposite is denoted with a '-'. When both a positive or a negative relation can be manifested depending on the properties of the process model, this is denoted with a '?' and when we were not able to identify the nature of the relation, no sign is represented on the relation.

## 3.4   Description of the existing process model quality structures (RQ 3)

Although we propose a new process model quality framework, it is useful to examine existing process model quality structures as well (RQ 3). We found four literature studies that have built a quality framework or model that integrates the knowledge about the different key questions (Q1-Q6) in a coherent whole.

### 3.4.1  The SIQ framework

The SIQ framework is a process model quality framework that tries to be simple (in contrast to for example the less user-friendly extensions of the SEQUAL framework) (Reijers et al., 2015). It has three concentric layers that jointly form a reference for making 'better' process models. From the inside to the outside we distinguish *the centre*, *the wall of checking* and *the wall of ensuring*. *The centre* contains a definition of process model quality based on the three quality dimensions of the SEQUAL framework: syntactic, semantic and pragmatic quality. The authors state that the syntactic quality forms the fundament, which supports the other two dimensions. Indeed, syntax errors can obscure the intended message of the process model, impacting the assessment of the semantic and pragmatic quality. The *wall of checking* layer bundles the initiatives to assess to which degree the three dimensions are valid. It is called *verification* of the syntactic quality, *validation* of the semantic quality and *certification* of the pragmatic quality. Similarly, the outer layer of the framework, *the wall of ensuring*, represents the initiatives to pro-actively enforce process model quality.

As is already mentioned by the authors, this framework is not representative for the whole domain knowledge at the time of its publication. A selection of a limited number of theories and methods was made based on their proven utility. Nevertheless, the authors are convinced that the framework has a great practical value due to its simplicity and integrative nature. They expressed the intention to complement the framework in future work with new insights and methods.

### 3.4.2  The BPMQ framework

Van Mersbergen (2013) presented a literature study that aimed at describing the *state of the art* of measuring and predicting process model quality and consolidating this knowledge in an orderly structure: the *Business Process Model Quality Framework*. This framework is visually composed of two pillars that each represent a different aspect of the quality research and that are connected by the relations between their components. In the right pillar, the three SEQUAL dimensions of process model quality are positioned: syntactic, semantic and pragmatic quality. They are the preferred dimensions for the 'WHAT' component of process model quality. They are complemented with operational definitions, associated with quality concepts, which are in turn related to the relevant metrics. In the left pillar, model-related metrics are placed that are considered to be predictors for process model quality, as well as a number of model characteristics that are related to the secondary notation of process models and which are not associated with metrics. The predictors of the left pillar are related to the quality metrics in the right pillar.

Van Mersbergen formulates a number of limitations. The author of this master thesis stresses that the choice to focus on the (limited) SEQUAL framework has an important impact on the scope of the BPMQ framework. The extensions of SEQUAL prove that there was a need for a more holistic view on process model quality and it may have been better to use one of these more evolved conceptual quality models as a starting point in order to capture the state of the art of the domain.

### 3.4.3 The MAQ model

Sadowska (2013) has built and validated a model for the assessment of the global quality of BPMN process models: i.e., the *Model for Assessing Quality of business process models in BPMN*. A meta-model was composed based on the ISO/IEC 1926 standard for software quality. This meta-model contains different entities: the quality model itself (MAQ), quality characteristics, quality sub-characteristics, quality functions, quality metrics and quality criteria with their composing scales. Eventually, a global quality function was developed for the MAQ, which is based on the weighted average of the quality functions of the sub-characteristics. A panel of BPMN experts determined the weights for this calculation by scoring the representativeness of the composing quality characteristics.

Two important remarks can be formulated about this approach. First, the author does not make a distinction between quality characteristics (dimensions) and impacting factors (drivers). For example, they include understandability as well as complexity as a quality characteristic. If a user-oriented approach is assumed (as in this paper, cf. Section 3.2.1), one can wonder to what degree it makes sense to include complexity as a quality characteristic. It can be argued that complexity in itself does not represent a fundamental user preference, but for example understandability, which is per definition a direct consequence of complexity, does. Note however, that this is a controversial opinion, as is illustrated by the great deal of disagreement about the 'WHAT' of process model quality. Second, the derivation of different quality categories for the model-related metrics is reported in a low-transparent way. It is not clear which quality dimensions were included in the statistical analysis in order to define the proposed thresholds, which seem to be determined by applying cluster analysis on the models of clearly 'high or low quality' (which was not explicitly defined). The global quality evaluation of the process models by the MAQ should thus be interpreted carefully.

### 3.4.4 The IQ-CC-EQ framework

Sánchez-González, et al. (2013) propose a model to objectively judge the quality of process models. As in the BPMQ framework, the quality assessment is enabled by the fact that the internal quality aspects of a process model can be measured and used as predictors for the related external quality dimensions. The internal quality (IQ) considers a process model as a white box and it targets the static properties of models that can be evaluated during the design phase. External quality (EQ) however, considers the model as a black box and targets those properties that relate to the impact of the model on its users. Both concepts are connected by the 'cognitive complexity' (CC), which is described by Briand et al. (1998, p. 349) as "*the mental burden [of a model] of the persons who have to deal with the class [this model]*". The internal quality influences the cognitive complexity, which influences in turn the external quality. Because the framework was not named, we refer to it as the IQ-CC-EQ framework, illustrating its composing concepts and their mutual relations.

The biggest concern with this framework is that the composing set of quality dimensions does not form a coherent whole. This is caused by the selection of the most popular characteristics of multiple standards and papers to build the framework. Although this way a certain consensus is reflected, a solid methodological basis, as in the conceptual quality frameworks such as SEQUAL, is lacking.

### 3.5 Evaluation of the existing process model quality structures (RQ 4)

As can be concluded from Table 2, each of the discussed quality structures of the previous section addresses a number of the defined key questions (Q1-Q6). Yet, none of them covers all aspects and not every covered aspect is completely addressed. The SIQ framework purposefully limits itself to a small number of trusted theories and methods, the BPMQ framework considers only three dimensions of quality, the MAQ targets only process models in BPMN and the IQ-CC-EQ framework focuses only on model-related drivers of process model quality. Furthermore, it can be noted how none of the structures contains answers to the theoretical 'WHY' question (Q6). One should however not forget that the goals of these structures are more specific than the goal of this paper. The discussed structures are thus not necessarily deficient, they just try to answer different research questions. Nevertheless, the research domain may benefit from a framework that tries to integrate the accumulated knowledge in a broad and comprehensive sense and to present it in an orderly manner.

Table 2. Evaluation of the existing process model quality frameworks

| Framework | Q1 | Q2 | Q3 | Q4 | Q5 | Q6 |
|---|---|---|---|---|---|---|
| SIQ | x | x | | | x | |
| BPMQ | x | x | x | x | | |
| MAQ | x | x | | | x | |
| IQ-CC-EQ | x | x | x | x | x | |
| CPMQF | x | x | x | x | x | x |

# 4 Discussion

Based on the developed framework, a number of observations related to the knowledge domain can be made. First, a number of authors evaluate process model quality explicitly by aggregating the quality dimensions with a certain fixed weighing factor to a single construct (e.g., Bandara and Gable, 2012; Sadowska, 2013). Although various studies stress out that the different quality dimensions should ideally be considered relatively to the intended goal of the process model (e.g., Bandara and Gable, 2012), these authors do not take the model goal into account. There were also no studies found that propose concrete goal-dependent weights, even though the need for these is clearly discussed in literature. This forms an interesting opportunity for further research.

Second, the framework shows that only a limited number of quality dimensions is measured. The research seems to focus on measuring syntactic correctness, understandability and modifiability. Moreover, the measurement of these last two dimensions corresponds with the measurement of the readability, the comprehensibility and the perceived readability of process models, which are all important to fulfil the communication goals of process models. No proxies or metrics were found for other quality dimensions. Is this caused by practical limitations, by disagreement about the relevance of certain dimensions or other reasons? Further research is required to answer these questions.

Third, for certain relations between quality drivers and directly measured quality dimensions a high number of studies was found that offer empirical proof for the significance of the proposed relations. For example, the significance for the relations between the model-related metrics number of nodes, structuredness and separability and the EPC-soundness criterion was confirmed by four different studies. This strongly supports the claim that these metrics can serve as predictors for the syntactic quality of process models. Nevertheless, insignificant results were not included in the framework and the quality of the primary studies was not investigated. Although they are out of the scope of this paper, in order to get a more elaborate view on the empirical validity of the relations, these critical investigations should be performed in the future.

Fourth, a number of model-related metrics stand out as validated predictors for the different dimensions of process model quality. For example, number of gateways, average gateway degree, maximum gateway degree, depth, control flow complexity, gateway heterogeneity and gateway mismatch show to have a significant relation with (proxies of) syntactical correctness, understandability, as well as modifiability. The evaluation of these metrics can thus have a positive impact on various dimensions of the quality of the process model. Nevertheless, the objection of the previous paragraph also applies here.

Finally, it was chosen to incorporate all the reported relations between the initiatives to realize process model quality and the drivers of process model quality into the framework. However, no studies were found that investigate the concrete impact of the implementation of certain initiatives on (the metrics of) the drivers. Therefore, the framework describes these potential relations without certainty that the reported relations for each initiative exist and are significant, nor that the set of displayed relations is complete. Due to a lack of empirical information about the influence of the initiatives on the drivers it is also not clear how exactly each initiative impacts the drivers, which forms an interesting opportunity for future research.

## 5    Conclusion

The main goal of this paper is the aggregation and structuring of the existing academic knowledge about process model quality, with a focus on completeness and relevance. In order to achieve this goal in an efficient manner, a tertiary literature study was executed. Via an automated search strategy, 3.056 unique studies were collected. Next, a two-phase selection process was applied on this collection, which resulted in a set of 42 relevant secondary studies. Based on a number of primitive interrogatives and the research goals of Genero et al. (2011), we defined six key questions to build a reference structure for the analysis. They are considered representative for the investigated aspects of process model quality. The analysis of a limited number of existing quality structures revealed that none of these integrates the knowledge about the six key questions or tackles all the addressed aspects entirely. In contrast, this paper explicitly pursues to build a framework that aggregates the collective knowledge and that structures it in order to raise the transparency of the research domain and to facilitate further research.

An important limitation of this literature review is that it is the result of the work of a very limited number of researchers. This threatens the validity because the selection, assessment, analysis and synthesis of the data can more easily be biased and be misinterpreted. Another crucial limitation of the paper is that the completeness of the presented information cannot be determined. Although this is typical for each literature study, this is important in this case because completeness is an explicit priority of the study. Therefore, a significant amount of consideration was spent on developing the search and selection strategies, which were also critically evaluated. The pursuit of completeness and relevance is a challenging balancing act between these two opposing goals. A third limitation relates to the uncertain or poor quality of the majority of the secondary studies. Those studies were not excluded in favour of completeness.

Future research could evaluate the quality of the information incorporated in the framework. Potentially, a number of relations should be omitted because the quality of the primary studies on which they are based can be questioned. The framework may also be complemented with new insights. It can be observed how the impact of personal and other drivers is not heavily investigated. A similar observation can be made about the impact of the initiatives to realize process model quality on the drivers of process model quality. The framework is deemed to be flexible enough to add such additional insights and aspects. However, because the interpretation of the framework in its current form could become (too) challenging, it seems valuable to investigate how the aggregated knowledge can be presented to different stakeholders with interactive data visualization techniques. Finally, we propose a thorough evaluation of the framework by other researchers in order to assess its true usefulness and the degree in which it achieves its goals of structuredness, completeness and relevance.

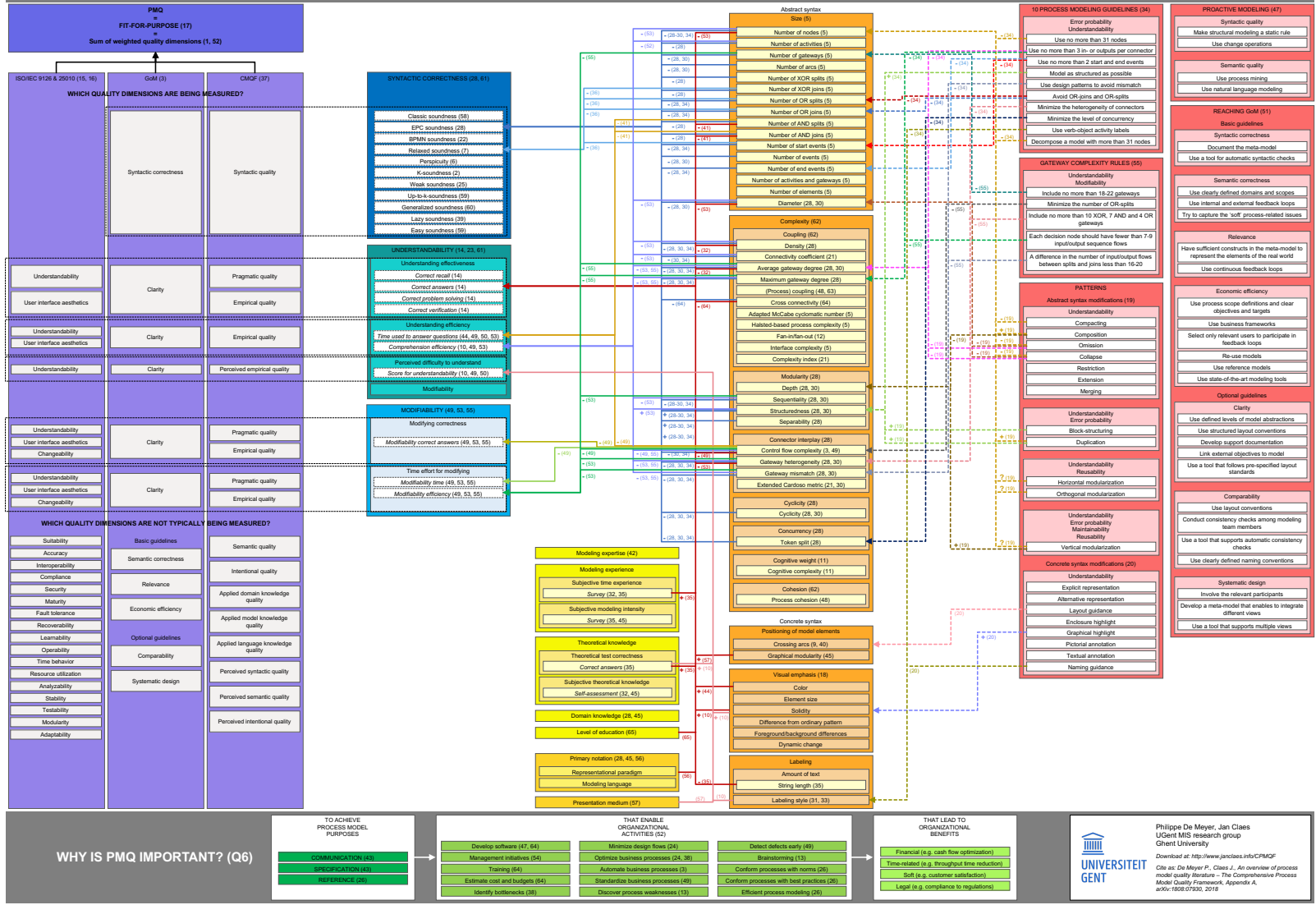

## LEGEND

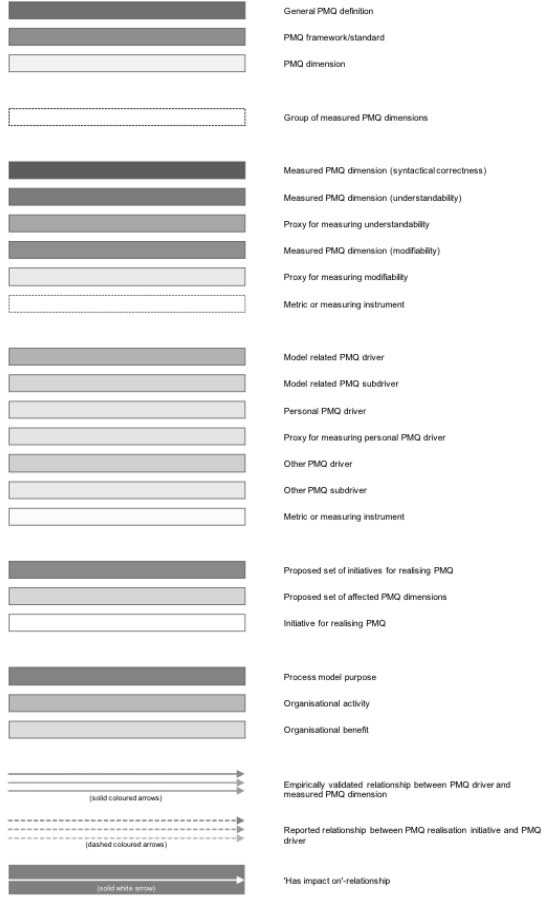